\documentstyle[aps,prb,epsf,twocolumn]{revtex}

\begin{document}

\title{Electron renormalization of sound interaction with two-level systems in
superconducting metglasses}

\author{E. V. Bezuglyi$^1$, A. L. Gaiduk$^1$, V. D. Fil$^1$, S. Zherlitsyn$^{1,3}$,
W. L. Johnson$^2$, G. Bruls$^3$, B. L\"uthi$^3$, and B. Wolf$^3$}

\address{$^1$B.Verkin Institute for Low Temperature Physics and Engineering,
47 Lenin Avenue, Kharkiv 310164, Ukraine\\
$^2$California Institute of Technology, Pasadena, CA 91125, USA\\
$^3$Physikalisches Institut, Universit\"at Frankfurt, Robert-Mayer-Str. 2-4,
60054 Frankfurt, Germany} \wideabs{ \maketitle \draft

\begin{abstract}
The crossing of temperature dependencies of sound velocity in the normal and
the superconducting state of metallic glasses indicates renormalization of
the intensity of sound interaction with two-level systems (TLS) caused by
their coupling with electrons. In this paper we have analyzed different
approaches to a quantitative description of renormalization using the results
of low-temperature ultrasonic investigation of
Zr$_{41.2}$Ti$_{13.8}$Cu$_{12.5}$Ni$_{10}$Be$_{22.5}$ amorphous alloy. It is
shown that the adiabatic renormalization of the coherent tunneling amplitude
can explain only part of the whole effect observed in the experiment. There
exists another mechanism of renormalization affecting only nearly symmetric
TLS.
\end{abstract}

 }

\section{introduction}

The well known tunneling model (TM) utilizes only two basic parameters for
the description of low-tem\-pe\-ra\-tu\-re ($T \lesssim 1$ K) behavior of the
velocity $v$ and the attenuation $\Gamma$ of sound in metallic
glasses.\cite{Hunk} A parameter $C_0=\overline p \gamma^2/\rho v^2$
[$\overline p$ is the density of states of two-level systems (TLS), $\gamma$
is the deformation potential, $\rho$ is the mass density] defines the scale
of variations of $v$ and $\Gamma$ in presence of TLS. A parameter $\eta=n_0
\sqrt{\overline {v_{kk^\prime}^2}}$ ($n_0$ is the density of electron states
at the Fermi level, $\overline{v_{kk^\prime}^2}$ is the mean square of matrix
element of electron-TLS scattering from $k$ to $k^\prime$ state) determines
the TLS relaxation rate due to their interaction with electron environment.
According to the TM, a TLS contribution to the acoustic characteristics is
determined by two additive mechanisms -- the resonance and the relaxation
ones. Under usual experimental conditions, $\omega \ll T$, where $\omega$ is
the sound frequency (we use the system of units where $\hbar =k_B=1$), the
resonance contribution to the variation $\delta v(T)$ of sound velocity is
always negative and represents a straight line with the unit slope in
coordinates $\delta v/C_0 v$ versus $\ln T$. The relaxation contribution is
also always negative and linear with the slope $-1/2$ in the same
coordinates. Thus, the resulting dependence $\delta v(\ln T)/C_0v$ is
expected to be a straight line with the slope $1/2$. The attenuation of sound
is determined mainly by the relaxation interaction and is virtually
independent on $T$, whereas the resonance contribution into $\Gamma$ is small
($\sim \omega /T$).

In superconducting glasses at low enough temperatures $T \ll T_c$ the
relaxation interaction is frozen out. This allows to extract purely resonance
contribution and therefore to verify many of TM conclusions. Acoustic
measurements in superconducting metglasses Pd$_{30}$Zr$_{70}$, \cite{Nec,Es}
Cu$_{30}$Zr$_{70}$,\cite{Esq} and
(Mo$_{1-x}$Ru$_x$)$_{0.8}$P$_{0.2}$,\cite{Lic} carried out more than a decade
ago, revealed some considerable deviations from the predictions of the TM:

i) the slope of the straight line $\delta v_n(\ln T)/C_0v$ in the normal
($n$) phase is about $1/4$ whereas the TM canonical slope is $1/2$;

ii) at least at high frequencies (HF), the normal state line $v_n(T)$ crosses
the superconducting ($s$) line $v_s(T)$ at $T_{\text{cr}}\ll T_c$. From the
TM point of view, this is impossible in principle;

iii) $v_s(T)$ is smaller than $v_n(T)$ just below $T_c$. This effect was
observed in both low frequency (LF) vibrating-reed
experiments\cite{Nec,Esq,Lic} and HF experiments.\cite{Es} According to the
original TM, the sound velocity would always increase below $T_c$;

iv) the sound attenuation reveals an analogous ano\-ma\-ly: $\Gamma_s(T)$
exceeds $\Gamma_n(T)$ within a certain temperature interval, which is about
$T/T_c \gtrsim 0.8$ in HF measurements\cite{Es} and extends down to $T/T_c
\gtrsim 0.05$ in LF experiments.\cite{Nec,Esq} In contrast, the TM predicts
the attenuation to be nearly independent on $T$ (with small $d\Gamma /dT>0$)
as long as the maximum relaxation rate $\nu$ exceeds $\omega$. Thus, the
attenuation in LF experiments should be insensitive to the superconducting
transition at all, whereas in HF measurements $\Gamma_s$ should either be
temperature-independent just below $T_c$ or decrease in metglasses with low
enough $T_c$ (or at high enough frequencies).

It was supposed in Refs.\ \onlinecite{Nec,Es} that all (or most of all)
deviations from the TM are related to the electron renormalization of the
parameter $C$ with respect to its bare value $C_0$. Although possible
mechanisms of this renormalization were not discussed in Refs.\
\onlinecite{Nec,Es}, the phenomenological consideration is rather simple.
Indeed, assume that $C$ decreases due to the interaction of TLS with the
electron excitations. As a result, the slope of $v_n(\ln T)$ decreases also.
On the other hand, the bare value $C_0$ should retrieve far below $T_c$.
Therefore, the ratio of slopes $v_n(\ln T)$ and $v_s(\ln T)$ becomes smaller
than the canonical TM value $1/2$. An additional assumption that the
parameter $C$ grows more rapidly just below $T_c$ than the relaxation
interaction is frozen out, leads to a simple explanation of the items iii)
and iv). A connection between the item ii) and the renormalization of $C$ is
less obvious. Nevertheless, we will demonstrate that the crossing between
$v_n(T)$ and $v_s(T)$ at ${T_{\text{cr}} \ll T_c}$ is the most convincing
evidence of reduced effective value of $C$ in the $n$-phase in comparison
with the $s$-phase.

The arguments in favor of the electron renormalization hypothesis have been
already presented in comparatively early theoretical works devoted to both
the general problem of tunneling with dissipation and a more detailed
analysis of the TLS interaction with surrounding electrons (see Refs.\
\onlinecite{Leg,Vla} and references therein). However, any relations allowing
to make a comparison with the experiment at a quantitative level were not
derived in these works.

For the first time, a straightforward theoretical analysis of the problem of
the electron renormalization of the sound-TLS interaction in metallic glasses
was made in Ref.\ \onlinecite{Kag}. It was argued that one of the reasons for
the decrease of $C$ in the presence of electrons is an adiabatic
renormalization of the coherent tunneling amplitude. Moreover, in order to
estimate this effect, it is not necessary to introduce any additional
parameter since the renormalization $(C_0-C)/C_0 = \eta^2 /4$ is determined
by the same interaction constant $\eta$ (see also Ref.\ \onlinecite{St}).
Although the theory \cite{Kag} gives some opportunity to examine its
conclusions quantitatively, such procedure was not accomplished, probably
because of lack of detailed experimental data.

In the present work we test different approaches to the quantitative analysis
of the sound velocity and attenuation in metglasses using experimental
results obtained on the superconducting amorphous
Zr$_{41.2}$Ti$_{13.8}$Cu$_{12.5}$Ni$_{10}$Be$_{22.5}$ alloy as an example. It
is shown that the adiabatic renormalization solely does not allow to describe
all the experimental results, and there exists an additional mechanism of
renormalization.

\section{Experimental results}

The alloy under investigation has a high resistance with respect to
crystallization in the state of overcooled melt and remains amorphous at
extremely low cooling rate ($< 10$ K s$^{-1}$).\cite{Pec} This makes it
possible to obtain bulk homogeneous samples, which suit perfectly the
acoustic measurements. The ultrasonic experimental technique is described
elsewhere.\cite{Lut}

Figure 1 shows typical temperature dependence of the velocity of the
transverse sound wave in Zr$_{41.2}$Ti$_{13.8}$
Cu$_{12.5}$Ni$_{10}$Be$_{22.5}$ in the $n$- and $s$-states. The $n$-state
measurements were carried out at the magnetic field $B = 1.5\div 2.5$ T. In
accordance with the TM, the curves $v(\ln T)$ represent almost straight lines
in both the $n$- and the deep $s$-state. The growth of $v_s$ below $ T_{c}$
reflects freezing out of the relaxation component and agrees with the TM
conception, with the constant $C_0=(2.85\pm 0.05)\cdot 10^{-5}$ determined
from the slope of $v_s(\ln T)$ at $T < 0.3$ K.

\begin{figure}[tbp]
\epsfxsize=8.5cm\epsffile{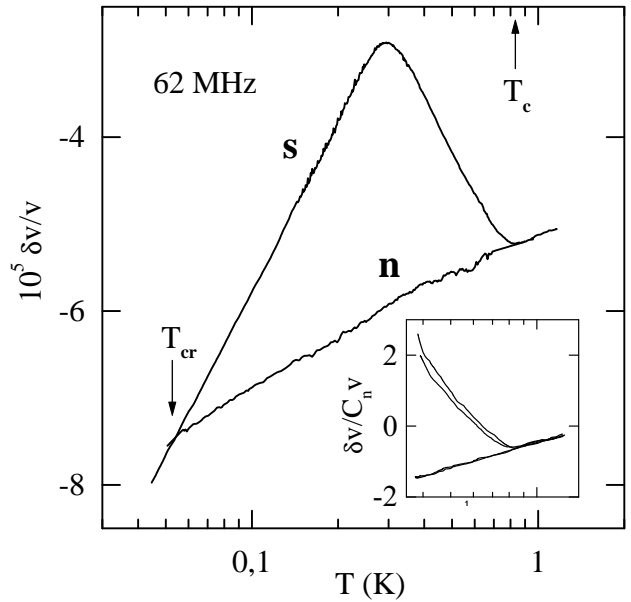} \vspace{0.2cm}
\caption{Temperature variations of the velocity of transverse sound in
Zr$_{41.2}$Ti$ _{13.8}$Cu$_{12.5}$Ni$_{10}$Be$_{22.5}$ alloy in the
superconducting and normal phases. Inset: normalized velocity of transverse
(upper trace, $C_n = 6.94\cdot 10^{-6}$) and longitudinal mode (lower trace,
$C_n = 2.75 \cdot 10^{-6}$) near $T_c$. The curves were aligned in the normal
phase.} \label{fig1}
\end{figure}

\begin{figure}[tbp]
\epsfxsize=8.5cm\epsffile{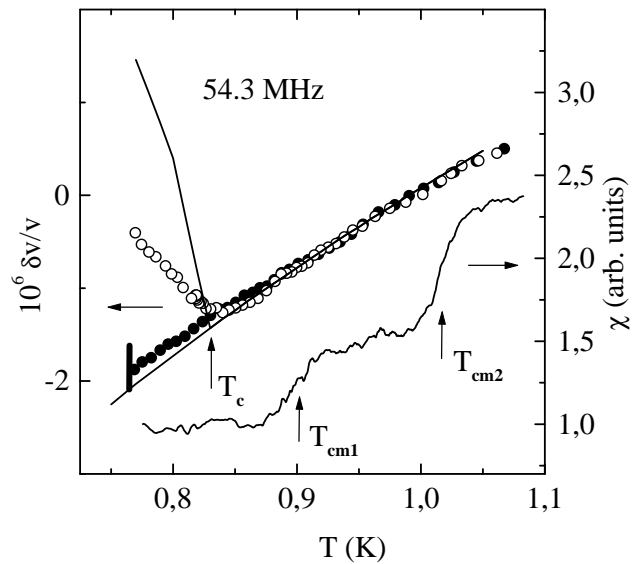} \vspace{0.2cm}
\caption{Temperature variations of magnetic susceptibility and sound velocity
in Zr$_{41.2}$Ti$_{13.8}$Cu$_{12.5}$Ni$_{10}$Be$_{22.5}$ in the vicinity of
$T_c$. Open and solid circles: $B=0$ and $B=1.5$ T. Solid lines: calculations
for $\eta=0.65$, $\varepsilon_b=1.2$, $u_b=0.5$, $R_s=0.14$, $T_c=0.83$ K.
Thick vertical mark shows noise level. Experimental data was smoothed by
adjacent averaging.} \label{fig2}
\end{figure}

There are also obvious deviations from the TM: the ratio of slopes in the
$n$- and the $s$-phases differs from its canonical value $1/2$ and is close
to $1/4$, and the curves $v(T)$ for both phases intersect at some temperature
$T_{\text{cr}}$. Such effects have been observed before in
Pd$_{30}$Zr$_{70}$.\cite{Es} The value of $T_{\text{cr}}$ is
frequency-dependent; particularly, $T_{\text{cr}}$(62 MHz) $\approx 0.055$ K
and $T_{\text{cr}}$(186 MHz) $\approx 0.11$ K were found.

The inset to Fig.\ 1 shows the variations $v(T)$ for transverse (t) and
longitudinal (l) sound normalized over correspondent slopes in the $n$-phase.
Obviously, these two dependencies virtually coincide, whereas absolute
velocity variations are sufficiently different.

Figure 2 shows temperature dependencies of $v_s$ and $v_n$ in the vicinity of
$T_c$ and the diamagnetic response of the sample on the ac magnetic field $H
=10^{-6}$ T at the frequency of 22 Hz. The magnetic susceptibility $\chi$ was
measured simultaneously with $v_s$ that provides coincidence of temperature
scales for both measurements. The presence of two steps in $\chi(T)$
indicates that the sample contains at least two phases with different
temperatures of the superconducting transition: $T_{\text{cm1}}\approx 0.9$ K
and $T_{\text{cm2}}\approx 1.0$ K. An increase of the ac field $H$ up to
$10^{-5}$ T leads to complete suppression of the anomaly in the diamagnetic
response at $T_{\text{cm2}}$, although the jump at $T_{\text{cm1}}$ survives
up to $H=10^{-4}$ T.

\begin{figure}[tbp]
\epsfxsize=8.5cm\epsffile{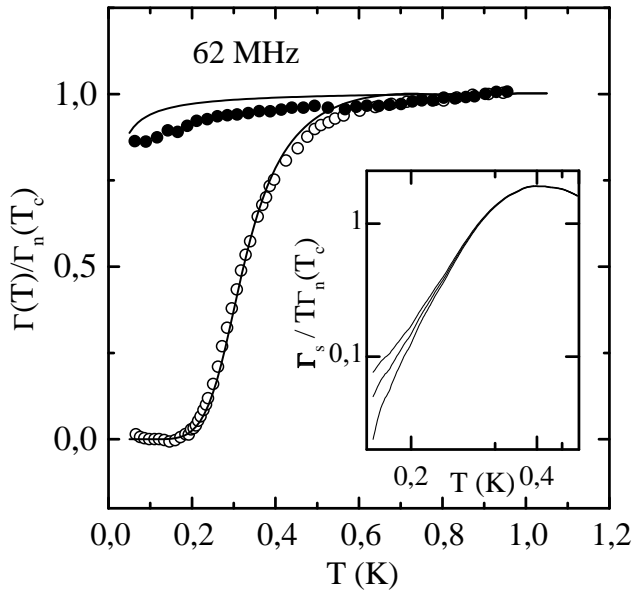} \vspace{0.2cm}
\caption{Normalized attenuation of the transverse sound wave versus
temperature in Zr$_{41.2}$Ti$_{13.8}$Cu$_{12.5}$Ni$_{10}$Be$_{22.5}$. Open
and solid circles: $B=0$ and $B=2.5$ T. Solid lines: calculated dependencies
for $\eta =0.65$, $T_c=0.83$ K. Inset: to determination of the
superconducting energy gap and the parameter $\eta$. The results are
presented for three values of the background level of attenuation: $-0.3\%$,
$0\%$ and $+0.3\%$ of the total change of attenuation between the $n$- and
$s$-phase, from bottom to top curve. } \label{fig3}
\end{figure}

It is of interest to note that the temperature $T_c\approx 0.83$ K, at which
one can first register a nonzero difference $v_s - v_n$, coincides neither
with $T_{\text{cm1}}$ nor with $T_{\text{cm2}}$. This was interpreted in
Ref.\ \onlinecite{Gai} as a fingerprint of possible gapless
superconductivity\cite{Ab} within the temperature region between $T_c$ and
$T_{\text{cm1}}$. It is known that the magnetic scattering, which is the most
prevalent reason of the gapless regime, also reduces the energy gap when the
latter opens. However, we found the energy gap in our alloy to be close to
the BCS value (see below) that allows us to reject this interpretation.
Apparently, the diamagnetic anomalies are related to some surface phases with
higher transition temperatures.

The temperature dependence of the sound attenuation is shown in Fig.\ 3. The
experimental data are normalized over $\Gamma_n(T_c)$ which can be determined
from the variations of sound amplitude between $T_c$ and deep $s$-state. The
behavior of $\Gamma(T)/\Gamma(T_c)$ does not show any noticeable difference
with similar dependencies in other superconducting amorphous alloys and
reflects evolution of the relaxation contribution to the attenuation, in
accordance with the TM conception.

\section{Qualitative consideration}

Before making quantitative estimations, we shall discuss qualitatively a
possible origin of the peculiarities of the sound velocity in superconducting
metglasses.

It is naturally to associate the crossing of $v_n(T)$ and $v_s(T)$ with a
growth of $C$ in the $s$-phase as a result of suppression of the electron
renormalization at energies smaller than the superconducting gap. To validate
this assumption, we address the expression for the resonant contribution of
TLS to the sound velocity: \cite{Pic}
\begin{equation}
\left({\delta v(T)\over v}\right)_{\text{res}}=P\int_0^\infty {CE\tanh(E/2T)
\over \omega^2-E^2}dE .
\eqnum{1}
\end{equation}

In the simple case of energy-independent $C = C_0$, in order to avoid a
formal logarithmic divergence of Eq.\ (1) at the upper limit, the velocity
variations are usually considered with respect to some arbitrary reference
temperature $T_0$:
\begin{equation}
\left(\displaystyle{v(T)-v(T_0)\over v(T_0)}\right)_{\text{res}} =C_0\ln
\displaystyle{T \over T_0}, \quad T>\omega.
\eqnum{2}
\end{equation}

However, if the value of $C$ varies with energy and/or temperature, Eq.\ (2)
is inapplicable even for qualitative estimates since the reference value
$v(T_0)$ may also change with $C$. To account correctly for the changes of
$C$, it is necessary to analyze the complete integral of Eq.\ (1) by
introducing a cutoff energy $E_m$ (say, of the order of melting temperature
or glass transition temperature). In the case of $C$ = const, Eq.\ (1) can be
approximated within the logarithmic accuracy by the following
piecewise-linear dependence (line 1 in Fig.\ 4):
\begin{equation}
\left( {\delta v(T)\over Cv}\right)_{\text{res}}=
\left\{\begin{array}{ccc}
\ln(\omega /E_{m}), & T \leq \omega \\ \ln(T/E_{m}), & T \geq \omega
\end{array}\right..
\eqnum{3}
\end{equation}
Here we neglect insignificant small variations in $v$ at $T\lesssim\omega$: a
quadratic fall near $T=0$ and a shallow minimum at $\omega=2.2T$ arising from
the analytical solution of Eq.\ (1).\cite{Hu}

\begin{figure}[tbp]
\epsfxsize=8.5cm\epsffile{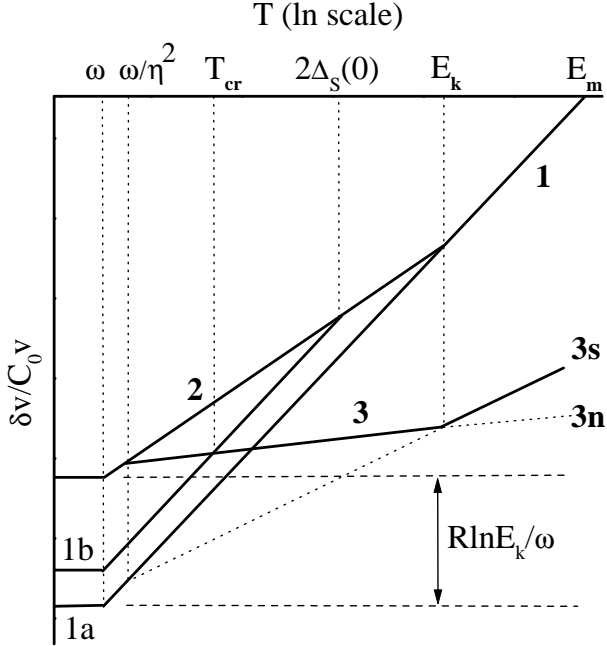} \vspace{0.2cm}
\caption{Schematic representation of temperature dependence of sound velocity
in superconducting glass. Lines 1, 1a and 1b correspond to $(\delta
v(T)/C_0v)_{\text{res}}$ in the superconducting phase. Line 2 represents
$(\delta v(T)/C_0v)_{\text{res}}$ in the normal phase. Line 3 is the total
$(\delta v(T)/C_0v)$ in the normal phase. See text for details.} \label{fig4}
\end{figure}

Since the parameter $C_0$ is determined for TLS in the absence of electron
excitations, the low-temperature part of line 1 in Fig.\ 4 depicts the
resonance contribution to the sound velocity at temperatures much lower than
$T_c$. Let us assume that the resonance interaction in the $n$-phase is
determined by a renormalized constant $C < C_0$ for all energies below a
certain threshold $E_k$. By making use of a simple step-like approximation
($C = C_0(1-R)$, $0 < R = \mbox{const} < 1$ for $E<E_k$ and $R=0$ in the
opposite case), one can plot the resonance contribution in the $n$-phase by
line 2 which is located over the corresponding line for the $s$-phase. Thus,
the resonance interaction increases the sound velocity in the $n$-phase at
$T<E_k$ if $C$ decreases and vice versa. This basic conclusion cannot be
derived from Eq.\ (2), where ``the dc component'' is lost.

The relaxation contribution to the sound velocity is much less sensitive to
the renormalization effects and becomes negligibly small at $T\lesssim \omega
/\eta^2$. This allows us to write down the following approximate expression:
\begin{equation}
\left( {\delta v(T)\over C_0v}\right)_{\text{rel}}= {1\over 2}
\left\{\begin{array}{ccc}
 0, & T\leq \omega /\eta^2 \\ \ln(\omega /T\eta^2), & T\geq \omega /\eta^2
\end{array}\right..
\eqnum{4}
\end{equation}

The total TLS contribution to the sound velocity in the $n$-phase is
schematically shown in Fig.\ 4 with a line 3 which was obtained in the
following way. At $T<E_k$ the line 3 contains a piece of line $3n$ drawn with
a slope $1/2-R$ through the point $T=\omega/\eta^2$ at the line 2. At $T
> E_k$, the line 3 is a piece of line $3s$ drawn with a slope $1/2$ from the
point $T=\omega/\eta^2$ at the line 1. A transition range arises at
temperatures $T\approx E_k$.

Let us now discuss the evolution of $v(T)$ at the superconducting transition.
First, we consider a low-tem\-pe\-ra\-tu\-re range $T\lesssim 0.3$ K. If $E_k
\leq 2\Delta_s(0)$ (here and below $\Delta_s(T)$ is the superconducting
energy gap), the renormalization of $C$ should be frozen out. Thus, the
resonance contribution can be depicted by a low-temperature part of the
straight line 1a. It is clear that the intersection of lines 3 and 1a is
possible only under the condition $C < C_0$ ($R \neq 0$). From the geometry
of Fig.\ 4 it is easy to estimate $T_{\text{cr}}$:
\begin{equation}
(1\!\!-\!S)\ln T_{\text{cr}}\! =\! R\ln E_k\! +\! (1\!\!-\!R\!-\!S) \ln
(\omega/\eta^2)\!\! -\! A(T_{\text{cr}}),
\eqnum{5}
\end{equation}
where $S$ is the resulting slope of $v_n(\ln T)$. The parameter $A$ is
introduced to account for possible shift of a background level of the sound
velocity in the $s$-phase with respect to the $n$-phase normalized over $C_0$
and will be discussed below. It can be seen from Eq.\ (5) that
$T_{\text{cr}}$ grows along with $E_k$, $\omega$, and with the decrease in
$\eta$.

At $E_k > 2\Delta_s(0)$ the renormalization of $C$ is only partially frozen
out for $E < 2\Delta_s(0)$. Therefore, the sound velocity in the deep
$s$-state can be depicted as a part of the straight line 1b (see Fig.\ 4).
The estimate of $T_{\text{cr}}$ by Eq.\ (5) is also valid in this case, only
$E_k$ should be replaced by $2\Delta_s(0)$.

Along this line of reasoning, we can also qualitatively explain the behavior
of $v$ and $\Gamma$ at the superconducting transition. Below $T_c$, the
electron renormalization rapidly reduces, and the effective $C$ grows
providing the decrease in $v$ and the increase in $\Gamma$. However, a
competitive effect arises simultaneously: the rate $\nu$ of the TLS
relaxation on electrons falls and therefore changes $v$ and $\Gamma$ in the
opposite direction. Thus, if the phonon relaxation predominates near $T_c$,
the effective $\nu$ changes weakly, and the sound velocity will decrease
(correspondingly, the attenuation will increase) below $T_c$, as it was
observed before.\cite{Nec,Es,Esq,Lic} If the electron relaxation prevails
(for materials with lower $T_c$ like our system), the changes of $v$ and
$\Gamma$ near $T_c$ may have any sign, depending upon the relations between
$T_c$, $E_k$ and $\omega$.

In principle, one can propose an alternative explanation of the crossing. In
our previous consideration, we silently assumed the count level of the TLS
contribution into the sound velocity to be the same for both the $n$- and
$s$-states. Generally, this is not the case, and the sound velocity changes
at the superconducting transition with no account of the TLS-related
mechanisms. For example, in pure metals a decrease in electron viscosity
below $T_c$ leads to the change of dislocation contribution to the sound
velocity of the order of $10^{-5}$ (Ref.\ \onlinecite{Fil}) which is
comparable with the TLS contribution but is undoubtedly absent in an
amorphous metal. A more general mechanism is the change of electron
contribution into elastic moduli of a metal in the $s$-phase. In disordered
metals with a short electron mean free path, this change is usually small
($\sim 10^{-6}$) but in certain cases, for instance, in A-15 compounds close
to structural instability, it may achieve much larger values $\sim 10^{-4}$
(Ref. \onlinecite{Tst}) of arbitrary sign. If we accept such a scale for the
decrease in the electron contribution in the $s$-state of our sample, the
crossing will arise without any renormalization effects. In its turn, the
anomalous slope ratio may be attributed to enhanced density of states of
asymmetric TLS playing the principal role in the sound attenuation. Although
the latter assumption contradicts the basic TM postulate about constancy of
$\overline{p}$ within a wide range of tunnel parameters, we can not reject
straight away the discussed alternative without additional argumentation
presented below.

Thermodynamic treatment \cite{Tst} shows that the electron contribution
variations in the $s$-state are independent on the sound frequency and lead
to a jump in derivatives $dv_i/dT$ at $T=T_c$ for both longitudinal and
transverse modes proportional to $\partial^2 T_c/ \partial e_i^2$ where
$e_{t,l}$ are corresponding deformations. A small jump of $v_l$ itself can be
also expected at $T=T_c$. As the temperature decreases, the electron
contribution changes as the density of a superfluid condensate, so that its
variations become negligibly small at $T \ll T_c$. By making use of Eq.\ (5)
at $R=0$ and the measured values of $S \approx 0.28$, $C_0$, $\eta \approx
0.65$ and $T_{\text{cr}}$(62 MHz), the corresponding shift of the sound
velocity between the $n$- and the deep $s$-state can be estimated as $\delta
v/v \sim 5 \cdot 10^{-5}$. This value is comparable with the resulting
velocity change $3\cdot 10^{-5}$ for t-mode (see Fig.\ 1) between $T_c$ and
the maximum in $v(T)$, i.e., the electron and the TLS contributions appear to
be of the same order but have opposite signs. However, since the normalized
TLS contribution is independent on the polarization, the data presented in
the inset in Fig.\ 1 demonstrate that the same independence should take place
for the electron contribution, i.e., the condition $(1/\gamma_l^2) \partial
^2 T_c/\partial e_l^2 \approx (1/\gamma_t^2) \partial ^2 T_c/ \partial e_t^2$
must be satisfied. The latter does not follow from theory and can be only a
result of random coincidence that is hardly possible. Thus, we conclude that
the scale of temperature variations of the TLS contribution much exceeds that
connected with the electron mechanisms. Furthermore, since the velocity shift
$A$ is frequency-independent, the following expression derived from Eq.\ (5)
\begin{equation}
\left(1-S \right) \ln {T_{\text{cr}}(\omega_1)\over T_{\text{cr}}(
\omega_2)}= (1-R-S) \ln {\omega_1 \over\omega_2}
\eqnum{6}
\end{equation}
shows that in the absence of renormalization ($R=0$) $T_{\text{cr}}$ must be
proportional to $\omega$ that contradicts our experimental data. In that way,
the absence of such proportionality is the most clear evidence of
renormalization of the parameter $C$ irrespectively of whether a certain
additional sound velocity shift between $n$- and $s$-phases exists or not.

\section{Some results of the tunneling model}

In this Section we present a brief overview of basic results of the TM, which
describe the behavior of $v(T)$ and $\Gamma(T)$ in glasses with account of
the dependence of $C$ on tunnel parameters and were used in our numerical
calculations. We also discuss modifications introduced into given relations
for a more exact account for the TLS-electron coupling.\cite{Kag,St}

The main postulate of the TM is a statement of the existence of double-well
potentials in glasses with the tunnel coupling between wells. The density of
states of TLS is constant in the space of parameters $\xi$, $\ln \Delta_0$
where $\xi$ is the asymmetry of the double-well potential and $\Delta_0$ is
an amplitude of the coherent tunneling. In order to determine the response of
the TLS ensemble on an external field, it is necessary to perform an
averaging over $\xi$ and $\ln\Delta_0$, which appears to be more convenient
in variables $E=\sqrt {\xi^2 + \Delta_0^2}$ and $u= \Delta_0 / E$. In this
representation, the TLS density of states is independent on $E$:
\begin{equation}
g(E,u)={\overline{p}\over u\sqrt{1-u^2}}\equiv g(u).
\eqnum{7}
\end{equation}

The relationships which determine the TLS contribution to the sound velocity
and attenuation read\cite{Hunk}:
\[ \left( {\delta v(T)\over v}\right)_{\text{res}} \!\!\! = \! -\!
\int_0^{E_m/T}\!\! \tanh \left({\varepsilon \over 2}\right)
{d\varepsilon\over\varepsilon} \int_0^1 C(\varepsilon,u)g(u) u^2du
\]
\begin{equation}
(\omega\ll T),
\eqnum{8}
\end{equation}
\[
\left({\delta v(T)\over v}\right)_{\text{rel}} =- {1\over 2}
\int_0^{E_m/T}{d\varepsilon\over\cosh^2(\varepsilon/2)} \int_0^1
C(\varepsilon,u) g(u)
\]
\begin{equation}
\times (1-u^2){\nu^2\over \omega^2+\nu^2}du,
\eqnum{9}
\end{equation}
\[
\left( {\Gamma v\over\omega}\right)_{\text{rel}}=\int_0^{E_m/T} {d\varepsilon
\over\cosh^2(\varepsilon / 2)} \int_0^1 C(\varepsilon,u) g(u)
\]
\begin{equation}
\times (1-u^2) {\omega\nu\over\omega^2+\nu^2}du.
\eqnum{10}
\end{equation}
In Eqs.\ (8)-(10) we introduced $\varepsilon = E/T$.

In general case, the relaxation rate $\nu$ is determined by both the
electrons and phonons, but for $T \lesssim 1$ K the phonon contribution can
be neglected. In the original TM the TLS-electron interaction is considered
within a perturbation theory\cite{Hunk} over the parameter $\eta^2$, which
does not affect the splitting of energy levels, and all specific features of
the metglass, in comparison with amorphous dielectrics, reduce only to the
appearance of a new relaxation channel having the rate
\begin{equation}
\nu={\pi \eta^2\over 2} u^2 T J(\varepsilon).
\eqnum{11}
\end{equation}

In the $n$-phase ${J(\varepsilon) = J_n(\varepsilon) = (\varepsilon/2)
\coth(\varepsilon /2)}$, $\nu \approx {\eta^2 T u^2}$, and the relaxation
interaction is essential for all $T > \omega$. In the $s$-state it is
necessary to use a function\cite{Bl}
\[
J_s(\varepsilon,\Delta)= {1\over 2}\int_\Delta^\infty d\varepsilon^\prime
{f(-\varepsilon^\prime)\over\sqrt{\varepsilon^{\prime 2}- \Delta^2}} \left\{
{\varepsilon^\prime(\varepsilon^\prime- \varepsilon)- \Delta^2 \over
\sqrt{(\varepsilon^\prime-\varepsilon)^2-\Delta^2}} \right.
\]
\begin{equation}
\times\!\! \left.{f(\varepsilon^\prime\!\!- \varepsilon)\over
f(-\varepsilon)} \Theta((\varepsilon^\prime\!\!-\varepsilon)^2\!\!- \Delta^2)
\text{sign} (\varepsilon^\prime\!\!-\varepsilon)+(\varepsilon \rightarrow\!\!
- \varepsilon) \right\}\!\!,
\eqnum{12}
\end{equation}
where $f(x)$ is the Fermi function, $\Theta(x)$ is the step Heaviside
function, and $\Delta = \Delta_s(T)/T$. This integral coincides with $J_n
(\varepsilon)$ for $\varepsilon \gg 2\Delta$, it has a discontinuity at
$\varepsilon = 2\Delta$, and  $J_s(\varepsilon,\Delta) \rightarrow 2f(\Delta
)$ for $\varepsilon \ll 2\Delta$. A rapid fall in $J_s$ below $T_c$ leads to
freezing out of the relaxation interaction when the maximal relaxation rate
($u=1$) becomes smaller than $\omega$.

A more complicated picture was revealed beyond the perturbation
theory.\cite{Kag,St} Just at $T=0$ in the $n$-phase, the bare amplitude of
the coherent tunneling $\Delta_0$ is renormalized due to an adiabatic part of
the interaction between the TLS and electrons:
\begin{equation}
\Delta_0^\ast\propto \Delta_0\left({\Delta_0\over\omega_0}\right)^
{\displaystyle {\eta^2\over 4-\eta^2}},
\eqnum{13}
\end{equation}
where $\omega_0$ is of the order of the Debye energy.

In the $n$-phase for $T\neq 0$ an ensemble of TLS can be divided into three
energy intervals:\cite{St}

1. $T \lesssim E^\ast =\sqrt{\xi^2+\Delta_0^{\ast 2}}$ -- coherently
tunneling TLS.

2. $E^\ast < T < 4\tilde{E}/\pi\eta^2$ -- incoherently tunneling TLS having
the tunneling amplitude
\begin{equation}
\widetilde{\Delta}\propto \Delta_0(2\pi T/\omega_0)^{\displaystyle \eta^2/4}
\eqnum{14}
\end{equation}
and the energy splitting of $\widetilde{E} = \sqrt{\xi^2+ {\widetilde
{\Delta}}^2}$. In new variables $E^\ast$, $\widetilde{E}$ and $u^\ast$,
$\widetilde{u}$ within the intervals 1, 2, respectively, the relations Eqs.\
(7)-(10) hold.

3. $T \gg 4\widetilde{E}/\pi\eta^2$ - low-energy TLS are relevant. In this
region, the amplitude of incoherent tunneling is also $\widetilde{\Delta}$ of
Eq.\ (14) but the factor $(1-{\widetilde{u}}^2)$ in Eqs.\ (9) and (10) is
absent because the incoherent transitions between the broadened levels take
place with energy variations even in the symmetric case. The corresponding
relaxation frequency varies as
\begin{equation}
\nu_3\propto{2\over\pi\eta^2}T{\widetilde{u}}^2 {\widetilde \varepsilon^2
\over J(\widetilde \varepsilon)}
\eqnum{15}
\end{equation}

One can think that a part of TLS with $\widetilde E < \sqrt{\omega T}$ should
decrease its contribution to $\Gamma$ and $(\delta v/v)_{\text{rel}}$ due to
the fall of $\nu_3$ for small $\widetilde{\varepsilon}$. However, a numerical
analysis shows that this fall is compensated by the growth of influence of
the symmetric TLS. As a result, partial contribution to $\Gamma$ and $(\delta
v/v)_{\text{rel}}$ from the interval 3 virtually does not change in
comparison with the original TM. The contribution of TLS from the second
interval remains the same also. Only the contribution from the coherently
tunneling TLS, undergoing the adiabatic renormalization, experiences an
essential change. The main postulate of the TM concerning the constancy of
$\overline p$ in the space of variables $\xi$, $\ln \Delta_0$ remains valid.
However, $g(u^\ast)$, along with the parameter $C$, acquires an additional
factor $(1-\eta^2/4)$ under transformation to the variables $E^\ast$,
$u^\ast$ because of a nonlinear dependence between $\Delta_0^\ast$ and
$\Delta_0$, Eq.\ (13). The latter changes rapidly at the superconducting
transition to the linear one of the type of Eq.\ (14), where $\Delta_s(T)$
substitutes for $T$, and all TM relations are restored.\cite{Kag}

\section{Analysis of the experimental data}

\subsection{Determination of $\Delta_s(0)$ and $\eta$}

For the frequencies used in our experiments, a rapid freezing of relaxation
interaction begins at the temperature well below $T_c$ (see Fig.\ 3). In this
case the renormalization of $C$ is also frozen out and the sound attenuation
$\Gamma_s$ is described by Eq.\ (10) with $C = C_0$. According to Eqs.\ (10)
and (11), the low-temperature part of $\Gamma_s(T)$ should be a straight line
in coordinates $\ln (\Gamma_s(T)/T)$ versus $T^{-1}$:
\begin{equation}
{\Gamma_s(T)\over\Gamma_n(T_c)}={2\pi\eta^2\over 3\omega}Te^{\displaystyle -
\Delta_s(0) /T}.
\eqnum{16}
\end{equation}

This allows us to use the sound attenuation for a simple evaluation of
$\Delta_s(0)$ and $\eta$ from its low-temperature dependence plotted in the
inset to Fig.\ 3. Since this construction is very sensitive to the reference
level of the attenuation, we also present two additional curves for level
variations of $\pm 0.3\%$ of the whole signal change between the $n$- and
$s$-states. Within this range, a large enough temperature interval exists
where each curve can be well approximated by a straight line with the slope
determining $\Delta_s(0) = 1.45\pm 0.05$ K. If we accept $T_c = 0.83$ K, this
value agrees well with the BCS relation $\Delta_s(0)/T_c = 1.76$ which was
used in all further calculations.

The value of $\eta = 0.55\pm 0.15$, determined by crossing of the
approximating straight lines with the ordinate axis for different curves in
the inset to Fig.\ 3, reveals a large spread due to its exponential
dependence on the position of the crossing point. A more accurate estimate of
$\eta$ can be obtained from the numerical analysis of the attenuation within
the whole temperature region of Fig.\ 3. By matching the most sharp part of
$\Gamma_s(T)$, calculated from Eq.\ (10) at $T_c = 0.83$ K, with the
experimental dependence, we found $\eta = 0.65 \pm 0.05$, in agreement with
previous rough estimate.

\subsection{Sound attenuation near $T_c$}

The analysis performed above shows that the temperature dependence of the
sound attenuation in Zr$_{41.2}$Ti$_{13.8}$Cu$_{12.5}$Ni$_{10}$Be$_{22.5}$
can be rather well described by the original TM. However, within some
temperature range just below $T_c$, the behavior of $\Gamma_s$ reveals
anomalies which find no explanation in the original TM.

\begin{figure}[tbp]
\epsfxsize=8.5cm\epsffile{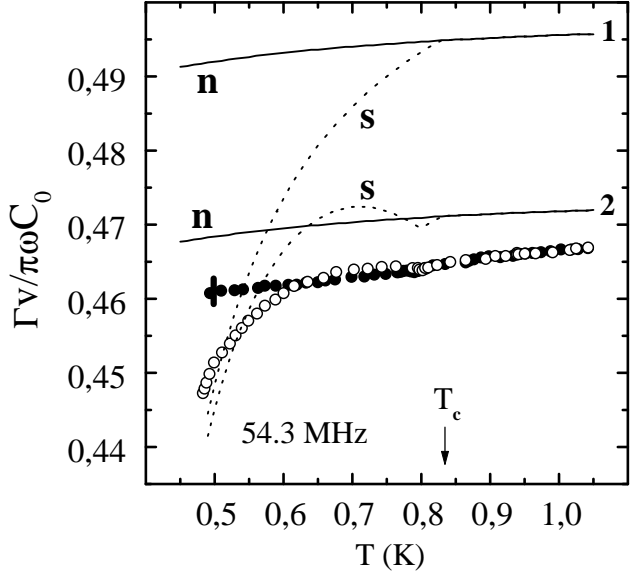} \vspace{0.2cm}
\caption{Comparison of sound attenuation near $T_c$ at 54 MHz with model
calculations. Open and solid circles: $B=0$ and $B=1.5$ T. Lines: related
calculation for the original TM (set 1 of lines) and the TM with account for
the adiabatic renormalization of the coherent tunneling amplitude, $\eta
=0.65$, $\varepsilon_b=1.2$ (set 2). Thick vertical mark shows noise level,
data was smoothed by adjacent averaging.} \label{fig5}
\end{figure}

Figure 5 shows variation of $\Gamma(T)$ in the vicinity of $T_c$ obtained
with higher resolution than in Fig.\ 3. The main peculiarities of
$\Gamma_s(T)$ can be discovered by means of a comparison of the experimental
curves with those calculated in the frame of the original TM (line set 1 in
Fig.\ 5). According to the calculation, a fall in the attenuation begins just
at $T_c$ with growing slope at low temperatures. The experimental dependence
has quite different behavior: $\Gamma_s(T)$ does not vary at $T_c$ within the
experimental resolution and tends to exceed $\Gamma_n(T)$ at lower
temperatures. A more prominent excess of $\Gamma_s(T)$ over $\Gamma_n(T)$
starting just at $T_c$ has been previously observed in Pd$_{30}$Zr$_{70}$
alloy.\cite{Es}

The proximity of $\Delta_s(0)/T_c$ to the BCS value, as well as the lowered
temperature of superconducting transition of the main volume of the sample in
comparison with $T_{\text{cm1}}$ (see Fig.\ 2) shows that the explanation of
$\Gamma_s(T)$ features by the magnetic depairing\cite{Gai} is irrelevant. Let
us now discuss the applicability of the concept of the electron
renormalization of $C$ to the description of the behavior of $\Gamma_s(T)$.

Indeed, Fig.\ 1 shows that the renormalization of $C$ really takes place that
follows from the crossing of $v_s(T)$ and $v_n(T)$ at $T_{\text{cr}} \ll
T_c$. However, the ratio of slopes of $v(\ln T)$ in the $n$- and $s$-states
points out that the scale of renormalization is rather large: $\delta
C/C_0\sim 0.22$, i.e., more than twice possible maximal contribution $\eta^2/
4\sim 0.1$ of the adiabatic renormalization of $\Delta_0$. Hence there should
be an additional mechanism of renormalization. Moreover, an incomparability
between the scale of $\delta C/C_0$ and the magnitude of the anomalies of
$\Gamma_s(T)$ indicates that this mechanism involves TLS which do not
contribute to the relaxation attenuation at $T\sim T_c$. Recall that the main
contribution to the attenuation comes from TLS with $\nu_{\text{opt}}\approx
\omega$ or $u_{\text{opt}}\approx \sqrt{\omega/ \eta^2T} \ll 1$.

One of possible additional mechanisms of the renormalization is associated
with the fluctuational modulation of a barrier in the double-well potential.
This mechanism affects only almost symmetric TLS ($u\sim 1$)\cite{Vla} and
does not give a contribution into $\Gamma_s(T)$.

According to Ref.\ \onlinecite{St}, the adiabatic renormalization of
$\Delta_0$ should involve all TLS with $E \gtrsim T$. This condition places
coherently tunneling TLS into the range where a cutting factor in the
denominator of Eq.\ (10) is relevant. In spite of that, their partial
contribution to $\Gamma(T)$ can be essential in the scale of Fig.\ 5.

In the original TM, the value of $\Gamma(T_c)$ for $\omega\ll T_c$ is close
to $1/2$ for the scale used in Fig.\ 5. The decrease in $C$ shifts
$\Gamma_n(T)$ towards smaller values. In order to analyze the shift of the
experimental dependence with respect to the calculated one, we need the
accuracy better than $1\%$ for the absolute value of the attenuation, which
is beyond the accuracy of our experiments. Therefore, in Fig.\ 5 we discuss
only the relative position of $\Gamma_n(T)$ and $\Gamma_s(T)$ curves, which
was measured much more precisely.

For numerical calculations we used the following energy dependence of the
parameter $C$:
\begin{equation}
{C\over C_0}=\!1\!-\!{\eta^2\over
4}\Theta(\varepsilon\!-\!\varepsilon_b)\left[ 1\!+\! (2f(\Delta)\!-\!1)\Theta
(2\Delta\!-\!\varepsilon)\right] \!-\!B.
\eqnum{17}
\end{equation}
This approximation is quite reasonable, since for $\varepsilon^\ast < 1$ the
coherent amplitude $\Delta_0^{\ast}$ decreases exponentially \cite{Kag}. Here
$\varepsilon_b \sim 1$ is a fitting parameter which confines the range of
coherently tunneling TLS. The last factor in the second term takes into
account that for $\varepsilon < 2{\Delta}$ the contribution comes only from
the normal excitations. The last term in Eq.\ (17) takes into consideration
an additional renormalization due to the symmetric TLS; its origin will be
discussed below. A contribution of $B$ to the sound attenuation is negligibly
small and at this stage we assume $B = 0$ for the sake of simplicity.

The results of simulation are plotted in Fig.\ 5 (lines 2). The difference
between lines 1n and 2n reflects a contribution from the adiabatic
renormalization of $\Delta_0$ to the sound attenuation in the $n$-state. We
matched almost temperature-independent part of $\Gamma_s(T)$ at $\eta = 0.65$
with measured curves and obtain quite reasonable value of $\varepsilon_b =
1.2 \pm 0.1$. The calculated dependence of $\Gamma_s(T)$ varies similarly to
the predictions of the original TM just below $T_c$. Then $\Gamma_s(T)$
undergoes a break, changing the sign of $d \Gamma/ dT$ at $T = 2 \Delta_s(T)/
\varepsilon_b$. These features arise due to exploiting a step approximation
in Eq.\ (17). The superconductivity does not affect the renormalization of
$C$ if $2 \Delta_s (T)$ is smaller than the value of $E = T \varepsilon_b$.
Obviously, a smoothed energy dependence of the cutoff factor in Eq.\ (17)
would decrease the variation of $\Gamma_s(T)$ at $T_c$ and eliminate the
break. The same result would also occur due to a possible broadening of the
superconducting transition in an amorphous sample.

In this way, the evolution of $\Gamma_s(T)$ in the vicinity of $T_c$ is
determined by two factors: a fall due to the decrease in the relaxation rate
$\nu$ and a growth because of freezing out of the $C$ renormalization. The
first factor is frequency-dependent in contrast to the second one. Therefore,
the resulting variations of $\Gamma_s$ should also depend on frequency. When
$\omega$ decreases, the temperature range where $\Gamma_s(T)>\Gamma_n(T)$
should be extended and vice versa. In particular, if $\eta$, $T_c$, and
$\varepsilon_b$ are fixed, the increase in frequency by an order of magnitude
(see, for example, Ref.\ \onlinecite{Es} where the measurement frequency was
about of 600 MHz) has to mask the action of the second factor utterly. This
frequency increase leads to $\Gamma_s(T)$ always lower than $\Gamma_n(T)$.
Besides, $d\Gamma_s(T)/dT$ grows as the temperature decreases. However, the
experiments in Ref.\ \onlinecite{Es} were carried out in metglass with
$T_c=2.6$ K where $\nu$ is essentially determined by phonons and depends
weakly on the state of electron subsystem. Under these circumstances, the
freezing out of the renormalization should give even stronger effect than
observed in our case.

\subsection{Sound velocity}

The resonance contribution to $v(T)$ is completely determined by coherently
tunneling TLS with $\varepsilon \gtrsim 1$. Therefore, the account of
adiabatic renormalization of $\Delta_0$ in the $n$-state will lead to
$(\delta v / C_0v)_{\text{res}} = (1- \eta^2/4) \ln T$ independently on the
magnitude of $\varepsilon_b$ in Eq.\ (17). From Eqs.\ (9) and (17), we
estimate a magnitude of the relaxation contribution as $(\delta v/ C_0
v)_{\text{rel}} = -1/2\left[1 - \eta^2/4(1-\tanh (\varepsilon_b/ 2))\right]
\ln T$. Using the values of $\eta$ and $\varepsilon_b$ obtained before, we
get $(\delta v/C_0 v)_n = 0.42\ln T$ for the total change in the $n$-state,
whereas the slope of the experimental dependence $(\delta v/C_0 v)_n =0.28\ln
T$ differs from the original TM coefficient $0.5$ much more. Thus, the
mechanism of the adiabatic renormalization of $\Delta_0$ can solely explain
less than a half of the whole effect, and, as it was mentioned above, an
additional origin of the electron renormalization of $C$ has to exist. It
must affect mainly the symmetric TLS which do not participate in the
relaxation attenuation. Besides, both mechanisms can be considered as
additive because the scale of $C$ renormalization is small.

The authors of Ref.\ \onlinecite{Vla} studied the effect of electron density
fluctuations on the barrier height of the interwell potential. It was argued
that below some critical temperature $T_k$, almost symmetric TLS and a
surrounding electron cloud can form a strongly-correlated (bound) state
similar to the Kondo state. This effect has an energy threshold, $E <
E_k(T_k)$. Fluctuations also lead to the renormalization of the tunnel
amplitude similar to Eq.\ (13):
\begin{equation}
\overline{\Delta_0}=\Delta_0 (T/D)^m,
\eqnum{18}
\end{equation}
where $D$ is of the order of the Fermi energy. The exponent $m$ depends on
$\eta$ and is close to $0.1\div 0.2$ for $\eta\sim 0.65$.

The spectrum transformation of Eq.\ (18) does not result in the
renormalization of the parameter $C$ since it does not change the density of
the TLS states within the space of new variables $\overline{u}$,
$\overline{E}$. These questions were not considered in Ref.\
\onlinecite{Vla}. Nevertheless, one can conclude that the renormalizing
factor would depend on $u=\Delta_0/E$ in Eq.\ (18) since, according to Ref.\
\onlinecite{Vla}, $\overline{\Delta_0}\rightarrow \Delta_0$ if $u\rightarrow
0$. This nonlinear relation between $\overline{\Delta_0}$ and $\Delta_0$
means the effective renormalization of TLS density of states similar to the
adiabatic renormalization. One can also expect a reduced value of the
deformation potential in the bound state.

Thus, the slope $(\delta v / C_0 v)_{\text{res}}$ is determined by the
relation $C/C_0=1-\eta^2 /4- R_s$ where $R_s$ describes the contribution from
symmetric TLS in n-phase. The latter does not change the slope of $(\delta v
/ C_0 v)_{\text{rel}}$, therefore $R_s$ is not a fitting parameter. Its value
for given $\eta$ is unambiguously determined by the resulting slope $S$ (with
the account for a small correction related to the contribution of adiabatic
renormalization of $\Delta_0$ in $(\delta v /C_0 v)_{\text{rel}}$).
Particularly, the estimates presented above give $R_s \approx 0.14$.

For computing, we model $B$ in Eq.\ (17) by simplest step function,
introducing a conventional boundary $u_b$ of the ``symmetric'' TLS:
\begin{equation}
B={R_s\over\sqrt{1\!-\!u_b^2}} \Theta (u\!-\!u_b)
[1\!+\!(2f(\Delta)\!-\!1)\Theta (2\Delta\! -\! \varepsilon )].
\eqnum{19}
\end{equation}

The meaning of two last factors in Eq.\ (19) is clear from preceding
discussion. The first factor represents the ``real'' renormalization by the
symmetric TLS because the renormalizing correction appears with the weight
$\sqrt{1- u_b^2}$ under the integration over $u$ in Eq.\ (8).

A comparison between calculated and experimental dependencies for $u_b=0.5$
is presented in Fig.\ 6. Here we use the following procedure. The
experimental points for the frequency of 62 MHz were taken from Fig.\ 1 and
normalized by $C_0$. It is impossible to measure mutual position of $v(T)$
for different frequencies with necessary accuracy of $10^{-7}$. However,
according to Eq.\ (3), the value $(\delta v(T)/v)_{\text{res}}$ does not
depend on frequency for $\omega \ll T$. That is why the experimental points
in the $s$-state for both frequencies 186 MHz and 62 MHz are placed together
at temperatures far below $T_c$. The positions of the calculated dependencies
are determined by the upper limit of integration in Eq.\ (7). Finally, one
can also compare them with experimental dependencies after matching of
$(\delta v(T)/Cv)_{\text{res}}$ in the $s$-state at $T\ll T_c$.

\begin{figure}[tbp]
\epsfxsize=8.5cm\epsffile{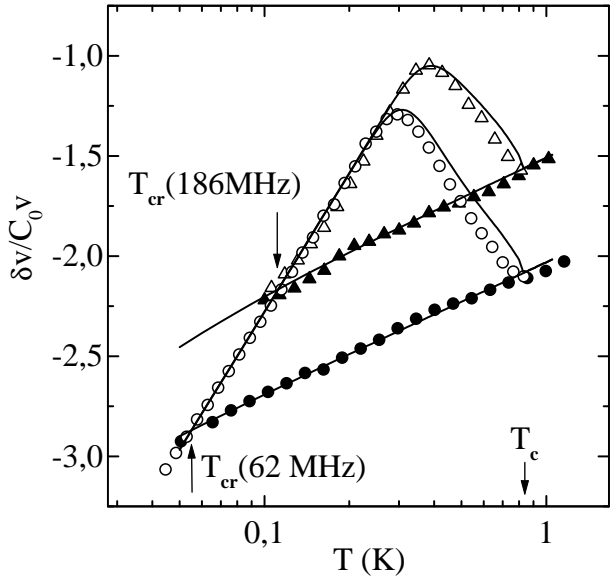} \vspace{0.2cm}
\caption{Comparison of the experimental change of sound velocity for
different frequencies and magnetic fields (circles: 62 MHz, triangles: 186
MHz; open symbols: $B=0$, solid symbols: $B=2.5$ T) with the calculations.
The following set of parameters is used: $\eta =0.65$, $\varepsilon_b=1.2$,
$u_b=0.5$, $R_s=0.14$, $T_c=0.83$ K.} \label{fig6}
\end{figure}

Figure 6 shows satisfactory agreement between the calculated and the
experimental dependencies. Some difference arises only in the temperature
range from $T_c$ to 0.4 K (see also Fig.\ 2). The calculated curve in the
$s$-phase is noticeably steeper at $T\sim T_c$ than the experimental one. An
estimate shows that a more smooth energy dependence of the adiabatic
renormalization of $\Delta_0$ leads only to insignificant decrease in the
slope of $(\delta v/C_0 v)_s$ at $T\sim T_c$. The most probable reason for
deviations of the computed dependencies from the experimental ones near $T_c$
is the smearing of superconducting transition and a contribution from small
thermodynamic corrections to $v(T)$ in the $s$-phase.\cite{Tst} A small
contribution can also come from a residual influence of phonon relaxation
which may cause the excess of the calculated slope of $(\delta v/C_0 v)_n$
over the experimental one near $T_c$ (Fig.\ 2).

The account for the symmetric TLS shifts up the resulting dependence $(\delta
v / C_0 v)_n$ at $R_s \ln 1.36 u_b$ without changing the slope of $(\delta
v/C_0 v)_{\text{rel}}$, as follows from computation of Eq.\ (9) with
renormalization of Eq.\ (19) for $\omega /\eta^2 T \ll u_b^2 \ll 1$. This
shift plays the same role as the parameter $A$ in Eq.\ (5). Therefore,
according to Eq.\ (5), the fitting of $u_b$ is reduced in fact to matching of
the computed $T_{\text{cr}}$ with the experimental value at given $\eta$,
$R_s$, and $S$. The fitted value of $u_b=0.5$ is smaller than $u_b\approx
0.7$ accepted in Ref.\ \onlinecite{Vla}. However, the latter is determined by
the bare magnitude of $\Delta_0$, whereas the integration in Eq.\ (8) is
performed over the renormalized variable $\overline{u}$. Therefore, according
to Eq.\ (18), $\overline{u_b}$ will be smaller than the bare value.

The account for the thermodynamic correction also will lead to an increase in
$u_b$; formally, $u_b$ can be made as much close to 1 as one likes.
Unfortunately, there is no clear way to separate the electron contribution on
the background of the TLS effects. Note that a weak temperature dependence in
Eq.\ (18) will introduce a correction into the slope $(\delta v(\ln T)/C_0
v)_{\text{rel}}$ too. Then the value $R_s$ in Eq.\ (19) should be decreased
by a factor of $(1+m)$.

Thus, our assumption about an additional mechanism renormalizing the
contribution of almost symmetric TLS to the sound velocity gives a
satisfactory description of the sound velocity behavior in both the $n$- and
$s$-states with a help of only one extra parameter $u_b$. Note that a
thorough calculation of $\Gamma_s(T)$ for $u_b=0.5$ with the additional
parameter $B$ in Eq.\ (16) leads to practically the same dependencies as
shown in Fig.\ 5 for $B=0$.

\section{Concluding remarks}

Using the results of the acoustic measurements obtained on the
superconducting glass Zr$_{41.2}$Ti$_{13.8}$Cu$_{12.5}$ Ni$_{10}$Be$_{22.5}$
we have carried out a quantitative analysis of different theoretical
approaches to the electron renormalization of sound interaction with TLS. A
convincing evidence for the renormalization is the crossing of the lines
$v_n(T)$ and $v_s(T)$ at some temperature $T_{\text{cr}} \ll T_c$, in
combination with the absence of proportionality between $T_{\text{cr}}$ and
$\omega$. Assuming the simplest model renormalization of the interaction
parameter $C$ in the space of tunnel variables, it is possible to describe
quantitatively the behavior of the sound velocity and attenuation exploiting
the original tunnel model. It is sufficient to use the adiabatic
renormalization of the coherent tunneling amplitude \cite{Kag,St} to fit the
dependence $\Gamma_s(T)$ with the experiment. However, the behavior of the
sound velocity can be described only with the help of an additional mechanism
of the renormalization affecting only almost symmetric TLS. This is the main
result of our consideration. The additional mechanism can be presumably
related to rebuilding of the interwell potential due to
fluctuations\cite{Vla} but this approach has not been developed enough to
consider it to be incontrovertible.

The analysis carried out in this work allows us to evaluate several
parameters using the experimental dependencies of $\Gamma(T)$ and $v(T)$. We
would like to emphasize that most of them are not fitting parameters in a
common sense because it is not necessary to vary the whole set of them
simultaneously to determine each of them. We utilize the following sequence
of parameter evaluation. First of all, from the temperature dependence of the
sound attenuation in the $s$-state we determine the energy gap $\Delta_s(0)$
and the parameter $\eta$. Using the latter together with the experimental
$\Gamma_s(T)$, $v_n(\ln T)$, and $T_{\text{cr}}$, we evaluate the parameters
$\varepsilon_b$, $R_s$, and $u_b$ consequently. In order to evaluate any
parameter, we utilize only the values which have been already found during
previous steps. In this way, $\eta$ is the central parameter which determines
all others. Therefore a question may rise: could an error in the
determination of $\eta$ lead to essential redistribution of the contributions
from different mechanisms? Under assumption that the electron contribution
into $v_s(T)$ does not change below $T_c$, the needed value of $u_b \sim 0.3$
is nonrealistically small already for $\eta=0.67\div 0.68$. On the other
hand, for $\eta\lesssim 0.6$ the best fit yields lower $T_c$. In this way,
the studied alloy does not give a large choice for the variation of the
parameters.

In principle, it would be possible to manage without the concept of
additional influence of electrons on the symmetric TLS in the analysis of
elastic properties of our metglass and reduce all effects to the adiabatic
renormalization of the tunnel amplitude, if we accept $\eta \approx 1$
assuming its preliminary estimate from low-temperature ``tail'' of the
attenuation (Sec.\ V, A) to be unreliable. However, in this case the overall
shape of calculated $\Gamma_s(T)$ can be adjusted to the experimental
dependence only at $T_c = 0.9$ K, revealing noticeable deviations from the
experimental data at $T\ll T_c$. Besides, the experimental value of
$T_{\text{cr}}$ could be obtained only under assumption that in $s$-state at
$T = T_{\text{cr}}$ the electron contribution is lowered by $\delta v/v
\approx 1.5\cdot 10^{-5}$ (this estimate follows from Eq.\ (5)) that is
comparable with the TLS contribution. Within this approach, the resulting
variations of $v(T)$ near $T_c$ would be determined by three equipollent
mechanisms: the change of the TLS relaxation rate, the freezing of the
adiabatic renormalization, and the evolution of electron contribution. In our
opinion, it is impossible to expect proper mutual compensation of their
partial contributions, for both the longitudinal and transverse modes,
providing regular variation of $v(T)$ observed at the superconducting
transition.

It is interesting to notice that the parameter $E_k$ which appears in the
fluctuation model\cite{Vla} is not so important in our evaluation process.
Our analysis does not demand also the absence of this parameter. Most
probably, the condition $E_k > 2\Delta_s(0)$ is satisfied in the alloy used
in our investigation because, as follows from Eq.\ (5), the introduction of
$E_k < 2\Delta_s(0)$ at constant $T_{\text{cr}}$ should be accompanied by a
decrease in $u_b$. So we believe that there is no need to use $E_k$ for the
description of low-temperature features of the sound velocity and attenuation
in our alloy. The introduction of $E_k$ can be probably more fruitful for the
analysis of the metglass elastic properties at higher temperatures.

\section{ Acknowledgments}

Authors thank G. Weiss for the fruitful discussion. This research was
partially supported by Ukrainian State Foundation for Fundamental Research
(Grant No2.4/153) and the Deutsche Forschungsgemeinschaft via SFB 252. W. L.
J. wishes to acknowledge the U. S. Dept. Of Energy for support under Grant No
DE-FG03-86ER45242. S. Z. would like to thank the Alexander von
Humboldt-Foundation for support.


\begin{references}
\bibitem{Hunk}
S. Hunklinger, and A. K. Raychaudhuri, {\it Progress in low temperature
physics}, edited by D. F. Brewer (North-Holland, Amsterdam, 1986), Vol.\ 9.

\bibitem{Nec}
H. Neckel, P. Esquinazi, G. Weiss, and S. Hunklinger, Sol. St. Comm. {\bf
57}, 151 (1986).

\bibitem{Es}
P. Esquinazi, H.-M. Ritter, H. Neckel, G. Weiss, and S. Hunklinger, Z. Phys.
B: Cond. Matter {\bf 64}, 81 (1986).

\bibitem{Esq}
P. Esquinazi, and J. Luzuriaga, Phys. Rev. B{\bf 37}, 7819 (1988).

\bibitem{Lic}
F. Lichtenberg, H. Raad, W. Moor, and G. Weiss, {\it Phonons 89}, edited by
S. Hunklinger, W. Ludwig, and G. Weiss (World Scientific, Singapore, 1990),
p. 471.

\bibitem{Leg}
A. J. Leggett, S. Chakravarty, A. T. Dorsey, M. P. A. Fisher, A. Garg, and W.
Zwerger, Rev. Mod. Phys. {\bf 59}, 1 (1987).

\bibitem{Vla}
K. Vl\'{a}dar, and A. Zawadowski, Phys. Rev. B{\bf 28}, 1564, 1582, 1596
(1983).

\bibitem{Kag}
Yu. Kagan, and N. V. Prokof'ev, Sol. St. Comm. {\bf 65}, 1385 (1988); Sov.
Phys. JETP {\bf 70}, 957 (1990).

\bibitem{St}
J. Stockburger, U. Weiss, and R. G\"orlich, Z. Phys. B: Cond. Matter {\bf
84}, 457 (1991).

\bibitem{Pec}
A. Pecker, and W. L. Johnson, Appl. Phys. Lett. {\bf 63}, 2342 (1993).

\bibitem{Lut}
B. L\"uthi, G. Bruls, P. Thalmeier, B. Wolf, D. Feinsterbusch, and L.
Kouroudis, J. Low Temp. Phys. {\bf 95}, 257 (1994).

\bibitem{Gai}
A. L. Gaiduk, E. V. Bezuglyi, V. D. Fil, and W. L. Johnson, Low Temp. Phys.
{\bf 23}, 857 (1997).

\bibitem{Ab}
A. A. Abrikosov and L. P. Gor'kov, Sov. Phys. JETP {\bf 12}, 1243 (1960).

\bibitem{Pic}
L. Pich\'e, R. Mayuard, S. Hunklinger, and J. J\"ackle, Phys. Rev. Lett. {\bf
32}, 1426 (1974).

\bibitem{Hu}
S. Hunklinger, and W. Arnold, {\it Physical Acoustics}, edited by W. P.
Mason, and R. N. Thurston, (Academic, New York, 1976), V.\ 12, p.\ 153.

\bibitem{Fil}
V. I. Denisenko, V. D. Fil, E. A. Masalitin, and P. A. Bezugly, Low Temp.
Phys. {\bf 7}, 529 (1981).

\bibitem{Tst}
L. R. Testardi, {\it Physical Acoustics}, edited by W. P. Mason, and R. N.
Thurston, (Academic, New York, 1973), V.\ 10, p.\ 193.

\bibitem{Bl}
J.L. Black, and P. Fulde, Phys. Rev. Lett. {\bf 43}, 453 (1979).
\end{references}
\end{document}